\documentclass[acmsmall]{acmart}
\usepackage{longtable}
\usepackage{enumitem}

\AtBeginDocument{%
  \providecommand\BibTeX{{%
    \normalfont B\kern-0.5em{\scshape i\kern-0.25em b}\kern-0.8em\TeX}}}

\setcopyright{rightsretained}
\acmJournal{PACMHCI}
\acmYear{2023} \acmVolume{7} \acmNumber{CSCW1} \acmArticle{145} \acmMonth{4} \acmPrice{}\acmDOI{10.1145/3579621}

\begin{document}

\title{Seeing Like a Toolkit: How Toolkits Envision the Work of AI Ethics}


\author{Richmond Y. Wong}
\email{rwong34@gatech.edu}
\orcid{0000-0001-8613-0380}
\affiliation{%
 \institution{Georgia Institute of Technology}
 \streetaddress{Digital Media}
 \city{Atlanta}
 \state{Georgia}
 \country{USA}
 \postcode{30308}
}

\author{Michael A. Madaio}
\email{michael.madaio@gmail.com}
\orcid{0000-0001-5772-0488}
\affiliation{%
 \institution{Microsoft Research}
 \city{New York City}
 \state{New York}
 \country{USA}
 \postcode{10012}
}

\author{Nick Merrill}
\email{ffff@berkeley.edu}
\orcid{0000-0003-3669-1387}
\affiliation{%
 \institution{University of California, Berkeley}
 \streetaddress{Center for Long-Term Cybersecurity}
 \city{Berkeley}
 \state{California}
 \country{USA}
 \postcode{94720}
}

\renewcommand{\shortauthors}{Wong, Madaio \& Merrill}

\begin{abstract}
    Numerous toolkits have been developed to support ethical AI development. However, toolkits, like all tools, encode assumptions in their design about what work should be done and how. In this paper, we conduct a qualitative analysis of 27 AI ethics toolkits to critically examine how the work of ethics is imagined and how it is supported by these toolkits. Specifically, we examine the discourses toolkits rely on when talking about ethical issues, who they imagine should do the work of ethics, and how they envision the work practices involved in addressing ethics. Among the toolkits, we identify a mismatch between the imagined work of ethics and the support the toolkits provide for doing that work. In particular, we identify a lack of guidance around how to navigate labor, organizational, and institutional power dynamics as they relate to performing ethical work. We use these omissions to chart future work for researchers and designers of AI ethics toolkits.
\end{abstract}

\ccsdesc[500]{Social and professional topics~Codes of ethics}
\ccsdesc[500]{Social and professional topics~Computing occupations}
\ccsdesc[300]{Social and professional topics~Computing organizations}

\keywords{fairness, ethics, toolkits, work, labor}


\maketitle

\section{Introduction}
\label{sec:intro}
Technology developers, researchers, policymakers, and others have identified the design and development process of artificial intelligence (AI) systems as a site for interventions to promote more ethical and just ends for AI systems \cite{Holstein:2019fr,madaio2020co,rakova2021responsible,schiff2020principles,madaio2022assessing}.
Recognizing this opportunity, researchers, practitioners, and activists have created a plethora of tools, resources, guides, and kits---of which the dominant paradigm is a ``toolkit'' \cite{lee2021landscape,richardson2021towards}---to promote ethics in AI design and development. Toolkits help technology practitioners and other stakeholders surface, discuss, or address ethical issues in their work. However, as the field appears to coalesce around this paradigm, it is critical to consider how these toolkits help to define and shape that work. Technologies that create standards (such as widely adopted toolkits), shape how people understand and interact with the world \cite{bowker1999sorting}

Prior research in CSCW and related fields has advanced our understanding of the work required to implement AI ethics principles in practice \cite[e.g.,][]{rakova2021responsible, madaio2020co,Holstein:2019fr,passi2019problem,madaio2022assessing}. 
In addition, prior work in CSCW has also examined the politics of tools and other artifacts designed to support the work of pursuing values and ethics \cite{Shilton2014HowToSeeValues,wong2020beyondchecklists}, such as security \cite{pierce2018differential}, privacy \cite{shilton2020rolepplaying,luger2015playing}, and UX design \cite[e.g.,][]{chivukula2021surveying}. 
Previous reviews of AI ethics and fairness toolkits have primarily focused on their usability and functionality \cite[e.g.,][]{lee2021landscape, richardson2021towards,deng2022exploring} or evaluating their efficacy in addressing ethical issues \cite[e.g.,][]{boyd2021datasheets}. In this paper, we contribute to these bodies of research by taking a more critical approach to understand how AI ethics toolkits, like all tools, enact values and assumptions about what it means to do the work of ethics. We start from the basis that simply creating toolkits will not be sufficient to address ethical issues. They must be adopted and used in practice within specific organizational contexts, but, as prior research has identified, adopting AI ethics tools and processes within organizational contexts presents challenges beyond usability and functionality  \cite[e.g.,][]{madaio2020co,rakova2021responsible,deng2022exploring,madaio2022assessing}. Therefore, by understanding how toolkits envision the work of AI ethics---particularly how those work practices may align (or not) with the organizational contexts in which they may be used---we may better identify opportunities to improve the design of toolkits and identify instances where additional processes or artifacts beyond toolkits may be useful. To investigate this, we ask: 

\begin{enumerate}[label=(RQ\arabic*)]
    \item What are the discourses of ethics that ethical AI toolkits draw on to legitimize their use? 
    \item Who do the toolkits imagine as doing the work of addressing ethics in AI? 
    \item What do toolkits imagine to be the specific work practices of addressing ethics in AI?
\end{enumerate}

To do this, we compiled and qualitatively coded a corpus of 27 AI ethics toolkits (broadly construed) to identify the discourses about ethics, the imagined users of the toolkits, and the work practices the toolkits envision and support. We found that AI ethics toolkits largely frame the work of AI ethics as technical work for individual technical practitioners, even as those same toolkits call for engaging broader sets of stakeholders to grapple with social aspects of AI ethics. In addition, we find that toolkits do not contend with the organizational, labor, and political implications of AI ethics work in practice.  In general, we found gaps between the types of stakeholders and work practices the toolkits call for and the support they provide. Despite framing ethics and fairness as sociotechnical issues that require diverse stakeholder involvement and engagement, many of the toolkits focused on technical approaches for individual technical practitioners to undertake. With few exceptions, toolkits lacked guidance on how to involve more diverse stakeholders or how to navigate organizational power dynamics when addressing AI ethics.\looseness=-1 


We provide recommendations for designers of AI ethics toolkits---both future and existing---to (1) embrace the non-technical
dimensions of AI ethics work; (2) support the work of engaging with
stakeholders\footnote{Here, we use the term ``stakeholder'' expansively, to include both potential users of the toolkits, others who may be part of the AI design, development, and deployment process, as well as other direct and indirect stakeholders who may be impacted by AI systems. We take this expansive approach following Lucy Suchman's work complicating the notion of the user \cite{suchman2002located} as well as Forlizzi and Zimmerman's work calling for more attention to stakeholders outside of the end users\cite{forlizzi2013promoting}. In cases where we specifically mean the users of the toolkit, we use the term ``user.''} from non-technical backgrounds; and (3) structure the work of AI
ethics as a problem for collective action. We end with a discussion of how we, as a research community, can foster the design of toolkits that achieve these goals, and we grapple with how we might
create metaphors and formats beyond toolkits that resist the 
solutionism\footnote{Although we provide suggestions for how to improve the design of AI ethics toolkits, we are wary of wholesale endorsing this form, as it may lead towards a technosolutionist approach. Nonetheless, this is the dominant paradigm for resources to support AI ethics in practice. As they are widely used, we believe there is value in exploring how the toolkit may be improved following the "practical turn" of values in design research \cite{flanagan2014values}, while simultaneously grappling with its limitations \cite[cf.][]{shilton2018values}.} prevalent in today's resources. 



\section{Background}

\subsection{Toolkits}

\subsubsection{As a genre}
What sort of thing is a toolkit? At their core, \textit{toolkits are curated collections of tools and materials}. Examples abound: do-it-yourself construction toolkits; first aid kits; traveling salesman kits; and research toolkits for (e.g.,) conducting participatory development efforts in rural communities \cite{mattern_2021, kelty_2018}, among many other examples. If we view them as a genre of communication \cite[cf.][]{yates1992genres}, we can see how their design choices structure their users' actions and interactions by conveying expectations for how they might be used. As Mattern has argued, toolkits make particular claims about the world through their design---they construct an imagined user, make an implicit argument about what forms of knowledge matter, and suggest visions for the way the world should be \cite{mattern_2021}. 
As a genre of communication, toolkits suggest a set of practices in a commonly recognized form; they formalize complex processes, but in so doing, they may flatten nuance and suggest that the tools to solve complex problems lie within the confines of the kit \cite{mattern_2021, kelty_2018}. Although artifacts can make certain practices legible, understandable, and knowable across different contexts, they can also abstract away from locally situated practices \cite{Scott1998seeing}. Moreover, toolkits work to configure what Goodwin calls professional vision: ``socially organized ways of seeing and understanding events that are answerable to the distinctive interests of a particular social group'' \cite[p606]{goodwin2015professional}.
This professional vision has political implications: in Goodwin's analysis, U.S. policing creates ``suspects'' to whom ``use of force'' can be applied \cite[p616]{goodwin2015professional}; it is thus critical to examine how toolkits may configure the professional vision of AI practitioners working on ethics.\looseness=-1



\subsubsection{In AI ethics}

In light of AI practitioners' needs for support in addressing the ethical dimensions of AI \cite{Holstein:2019fr}, technology companies, researchers at CSCW, FAccT, CHI, and other venues, as well as other groups have developed numerous tools and resources to support that work, with many such resources taking the form of toolkits \cite[e.g.,][]{lee2021landscape,richardson2021towards,morley2021initial,10.1145/3442188.3445938,shen2021valuecards,gebru2021datasheets,mitchell2019model,deng2022exploring, shen2022model}. 
Several papers have performed systemic meta-reviews and empirical analyses of AI ethics toolkits \cite{lee2021landscape,morley2021initial,richardson2021towards,ayling2021putting,deng2022exploring}. For instance, one line of research performs descriptive analyses of AI ethics toolkits, including \citet{ayling2021putting}'s work identifying stakeholder types common across toolkits, and stages in the organizational lifecycle at which various toolkits are applied, as well as \citet{morley2021initial}'s work proposing a typology of AI ethics approaches synthesized from a variety of toolkits, and \citet{Crockett2021BuildingTrustworthy}'s analysis of 77 AI ethics toolkits, finding that many lack instructions or training to facilitate adoption. 
In addition, others have conducted more empirical examination of toolkits, including \citet{lee2021landscape}'s normative evaluation of six open source fairness toolkits, using surveys and interviews with practitioners to understand the strengths and weaknesses of these tools, as well as \citet{richardson2021towards}'s work conducting simulated ethics scenarios with ML practitioners, observing their experience using various ethics toolkits to inform recommendations for their design, and \citet{deng2022exploring}'s work exploring how practitioners use toolkits in their AI ethics work in practice. 

In technology fields other than AI ethics, others have studied how design toolkits shape work practices.
For instance, \citet{chivukula2021surveying} identify how toolkits operationalize ethics, identify their audience, and embody specific theories of change. 
\citet{pierce2018differential}'s analysis of cybersecurity toolkits reveals a complex set of ``differentially'' vulnerable persons, all attempting to achieve security for their socially situated needs. Building on prior empirical work evaluating the functionality and usability of AI ethics toolkits, we take a critical approach to understand the \textit{work practices} that toolkits envision for their imagined users, and how those work practices might be enacted in particular sites of technology production.
In other words, we focus our analysis on how toolkits help configure the \textit{organizational practice} of AI ethics. 


\subsection{AI Ethics in Organizational Practice}
As the field of AI ethics has moved from developing high-level principles \cite{Jobin:2019bw} to operationalizing those principles in particular sets of practices \cite{Mittelstadt:2019ve, schiff2020principles}, prior research has identified the crucial role that social and organizational dynamics play in whether and how those practices are enacted in the organizational contexts where AI systems are developed \cite{Metcalf2019OwningEthics, madaio2020co, rakova2021responsible}. Substantial prior work has identified the crucial role of organizational dynamics (e.g., workplace politics, institutional norms, organizational culture)  in shaping technology design practices more broadly \cite{suchman2002located, wong2021tactics, shilton2013values, neff2020bad}. Prior ethnographic research on the work practices of data scientists has identified how technical decisions are never just technical---that they are often contested and negotiated by multiple actors (e.g., data scientists, business team members, user researchers) within their situated contexts of work \cite{passi2019problem,passi2018trust}. \citet{passi2020making} discuss how such negotiations were shaped by the organizations' business priorities, and how the culture and structure of those organizations legitimized technical knowledge over other types of knowledge and expertise, in ways that shaped how negotiations for technical design decisions were resolved. These dynamics are found across a range of technology practitioners, including user experience professionals \cite{wong2021tactics, Chivukula2020DimensionsUX}, technical researchers \cite{shilton2013values}, or privacy professionals \cite{Bamberger2015PrivacyGround}.


Prior research on AI ethics work practices has similarly identified how the organizational contexts of AI development shape practitioners' practices for addressing ethical concerns. Metcalf et al., explored the recent institutionalization of ethics in tech companies by tracing the roles and responsibilities of so-called ``ethics owners'' \cite{Metcalf2019OwningEthics}. In contrast with ethics owners who may have responsibility over ethical implications of AI, \citet{madaio2020co} identified how the social pressures on AI practitioners (e.g., data scientists, ML engineers, AI product managers) to ship products on rapid timelines disincentivized them to raise concerns about potential ethical issues. Taking a wider view, \citet{rakova2021responsible} discussed how AI development suffers from misaligned incentives and a lack of organizational accountability structures to support proactive anticipation of and work to address ethical AI issues. However, as resources to support AI ethics work have proliferated---including AI ethics toolkits---it is not clear to what extent the designers of those resources have learned the lessons of this research on how organizational dynamics may shape AI ethics work in practice.

\section{Methods}
\subsection{Researchers' positionality}

The three authors share an interest in issues related to fairness and ethics in AI and ML systems, and have formal training in human-computer interaction and information studies, but also draw on interdisciplinary research fields studying the intersections of technology and society. All three authors are male, and live and work for academic and industry research institutions in the United States.
One author's 
prior research is situated in values in design, studying the practices used by user experience and other technology professionals to address ethical issues in their work, including the organizational power dynamics involved in these practices. 
Another author's prior work has focused on how AI practitioners conceptualize fairness and address it in their work practices. He has conducted fairness research with AI practitioners, has contributed to 
multiple resources for fairness in AI, and has worked on fairness in AI at large technology companies.
The third author 
has built course materials to teach undergraduate and graduate students how to identify and ameliorate bias in machine learning algorithms and has reflected on the 
ways that students do not get exposed to fairness in technical detail during their coursework. 
\looseness=-1 

The corpus we developed may have been shaped by our positionality as researchers in academia and industry living in the U.S. and conducting the search in English. Our prior research with technology practitioners led us to focus on the artifact of the ``toolkit,'' which we have encountered in our prior work, although we recognize that this focus may obscure other artifacts and forms of action that are currently in use but that did not fit our conception of a toolkit. Furthermore, our familiarity with gaps between the corporate rhetoric of ethical action and actual practices related to ethical action (e.g., \cite{Hoffmann2020terms}) led us to focus our research questions and analysis to highlight potential gaps between the rhetoric or imaginaries embedded in toolkits and the practices or tensions we are familiar with from our prior work and experiences with practitioners. This framing is one particular lens with which to understand these artifacts, although there may be other lenses that may provide additional insights.

\subsection{Corpus development}
We conducted a review of existing ethics toolkits, curated to explore the breadth of ways that ethical issues are portrayed in relation to developing AI systems. We began by conducting a broad search for such artifacts in May-June 2021. We searched in two ways. First, we looked at references from recent research papers from CSCW, FAccT, and CHI that survey ethical toolkits \cite[e.g.,][]{lee2021landscape,richardson2021towards}. Second, following the approach in \citet{lee2021landscape}, we emulated the position of a practitioner looking for ethical toolkits and conducted a range of Google searches for artifacts using the terms: ``AI ethics toolkit,'' ``AI values toolkit,'' ``AI fairness toolkit,'' ``ethics design toolkit,'' ``values design toolkit.'' Several search results provided artifacts such as blog posts or lists of other toolkits, and many toolkits appeared in results from multiple search terms.\footnote{Although not all toolkits specifically focused on AI (some focused on ``algorithms'' or ``design''), their content and their inclusion in search results made it reasonably likely that a practitioner would consult with the resource in deciding how to enact AI ethics.}
We shared and discussed these resources with each other to discuss what might (not) be considered a toolkit (for instance, we decided to exclude ethical oaths or compilations of tools).\footnote{Note that the term \textit{toolkit} is used in this paper is an analytical category chosen by the researchers to search for and describe the artifacts being studied. Not all the artifacts we analyzed explicitly described themselves using the term toolkit. See the Appendix for more details about the toolkits.}
Although we broadly view toolkits as curated collections of tools and materials, we largely take an inductive approach to understanding what toolkits purport to be.
From these search processes, we initially identified 57 unique candidate toolkits for analysis.

Our goal was to identify a subset of toolkits for deeper qualitative analysis in order to sample a variety of types of toolkits (rather than attempt to create an exhaustive or statistically representative sample). After reading through the toolkits, we discussed potential dimensions of variation, including: the source(s) of the toolkit (e.g., academia, industry, etc), the intended audience or user, form factor(s) of the toolkit and any guidance it provided (e.g., code, research papers, documentation, case studies, activity instructions, etc.), and its stated goal(s) or purpose(s).
 %
 We also used the following criteria to narrow the corpus for deeper qualitative analysis: 
 \begin{itemize}
     \item \textit{The toolkit's audience should be a stakeholder related to the design, deployment, or use of AI systems.} This led us to exclude toolkits such as Shen et al.'s value cards \cite{shen2021valuecards}, designed primarily for use in a student or educational setting, but \textit{not} to exclude toolkits such as \citet{10.1145/3442188.3445938}, intended to be used by community advocates. 
        We excluded five artifacts that focused on non-AI systems, and four designed to be used in classroom settings.
        \looseness=-1 
     \item \textit{The toolkit should provide specific guidance or actionable items to its audience}, which could be technical, organizational, or social actions. Artifacts that provided lists of other toolkits or only provided informational materials were excluded (e.g., a blog post advocating for greater use of value-sensitive design \cite{Shonhiwa2020humanValuesMedium}). 
        We excluded five artifacts that were primarily informational or advocacy materials, four where we could not access enough information, such as paywalled services, and two that focused on professional education activities.
     \item \textit{Given our focus on practice, the toolkit should have some indication of use} (by stakeolders either internal or external to companies). Although we are unable to validate the extent to which each toolkit has been adopted, 
     we used a set of proxies to estimate which toolkits are likely to have been used by practitioners, including whether it appeared in practitioner-created lists of resources, its search results rankings, or (for open source code toolkits) indications of community use or contributions. One author also works in an industry institution, and was able to provide further insight into toolkit usage by industry teams. This excluded some toolkits that were created as part of academic papers, and which did not seem to be more broadly used by practitioners at the time of sampling, such as FairSight \cite{ahn2020fairsight}.  
        We excluded seven artifacts that seemed to have low use, and two artifacts that were primarily academic research papers.
    \item In addition, due to the authors' language limitations, we excluded one toolkit not in English.
 \end{itemize}
 
We independently reviewed the toolkits for inclusion, exclusion, or discussion. As a group, we discussed toolkits that we either marked for discussion or that we rated differently. To resolve disagreements, we decided to aim for variation along multiple dimensions (a toolkit that overlapped a lot with an already included toolkit was less likely to be included). From the 57 candidates, 30 total were excluded.
The final corpus includes 27 toolkits, which are summarized in Section \ref{section:corpus-description} and fully listed in Appendix \ref{section:toolkit-list}. 

\subsection{Corpus Analysis}
In the first round of our analysis, we conducted an initial coding 
of the 27 toolkits based on the following dimensions: the source(s) of the toolkit (e.g., academia or industry), the intended audience or user, its stated goal(s), and references to the ML pipeline.\footnote{Although many of these were explicitly stated in the toolkits' documentation, some required some interpretative coding. We resolved all disagreements through discussion amongst all three authors.} We used the results of this initial coding to inform our discussions of which toolkits to include in the corpus, as well as to inform our second round of analysis. We then began a second round of more open-ended inductive qualitative analysis based on our research questions (following \cite{braun2006using}). From reading through the toolkits, the authors discussed potential emerging themes. 
These initial themes included: what work do toolkits imagine is needed to address AI ethics; who do toolkits describe as doing the work of AI ethics; how does that compare to prior research about enacting AI ethics work in practice; what types of guidance are provided in toolkits; how do toolkits refer to the organizational contexts where they may be used; how do toolkits conceptualize social values (such as fairness or inclusion); when in or beyond the design process do the toolkits suggest they should be used; the toolkits' different form factors; what social or technical background knowledge might be required to understand or use the toolkit; and whether toolkits describe any risks or limitations associated with their use. Our open-ended exploration of these themes helped us refine our research questions (to those presented in Section \ref{sec:intro}).

Based on these themes, we decided to ask the following questions of each of the toolkits to further our analysis:
\begin{itemize}
    \item What language does the toolkit use to describe values and ethics?
    \item What does the toolkit say about the users and other stakeholders of the AI systems to whom the toolkit aims its attention?
    \item What type of work is needed to enact the toolkit's guidance in practice?
    \item What does the toolkit say about the organizational context in which workers must apply the toolkit?
\end{itemize}

Each author read closely through one third of the toolkits, found textual examples that addressed each of these questions, and posted those examples onto sticky notes in an online whiteboard. Collectively, all the authors conducted thematic analysis and affinity diagramming on the online whiteboard, inductively clustering examples into higher-level themes, which we report on in the findings section. 

\subsection{Corpus Description}
\label{section:corpus-description}


We briefly describe our corpus of 27 toolkits based on our first round of analysis.\footnote{Multiple codes could be assigned to each toolkit, so the counts may sum to more than 27.}
A full listing of toolkits is in Appendix \ref{section:toolkit-list}, including details of our coding results in Table \ref{tab:toolkits-analysis}. 
The \textbf{toolkit authors} include: technology companies (16 toolkits), university centers and academic researchers (6), non-profit organizations or institutes (6), open source communities (2), design agencies (2), a government agency (1), and an individual tech worker (1). 

The toolkits' \textbf{form factors} vary greatly as well. Many are technical in nature, such as open-source code (11 toolkits), proprietary code (1), documentation (12), tutorials (2), a software product (1), or a web-based tool (1). Other common forms include exercise or activity instructions (7), worksheets (5), guides or manuals (5), frameworks or guidelines (2), checklists (2), or cards (2). Several include informational websites or reading materials (4). 

Considering the toolkits' \textbf{audiences}, most are targeted towards technical audiences such as developers (6 toolkits), data scientists (6), designers (5), technology professionals or builders (3), implementation or product teams (3), analysts (2), or UX teams (1). Some are aimed at different levels within organizations, including: managers or product/project managers (2), executive leadership (1), internal stakeholders (1), team members (1), or organizations broadly (1). Some toolkits' audiences include people outside of technology companies, including: policymakers or government leaders (3), advocates (3), software clients or customers (1), vendors (1), civil society organizations (1), community groups (1), and users (1). We elaborate more on the toolkits' intended audiences in Section \ref{stakeholders}.\looseness=-1

\section{Findings}

We begin our findings with a description of the language toolkits use to describe and frame the work of AI ethics (RQ1). We then discuss the audiences envisioned to use the toolkits (RQ2); and close with what the toolkits envision to be the work of AI ethics (RQ3). 

\subsection{Language, framing, and discourses of ethics (RQ1)}
\label{discourses}



\subsubsection{Motivating Ethics: Harms, Risks, Opportunities, and Scale}
We first look at how the toolkits motivate their use. Often, they articulate a problem that the toolkit will help address.
One way of articulating a problem is identifying how AI systems can \textbf{have effects that harm people.} 
In such cases, toolkits motivate ethical problems by highlighting harms to people outside the design and development process---a group that Pfaffenberger terms the ``impact constituency,'' the ``individuals, groups, and institutions who lose as a technology diffuses throughout society'' \cite[p297]{Pfaffenberger1992}. 
For instance, Fairlearn describes unfairness ``in terms of its impact on people — i.e., in terms of harms — and not in terms of specific causes, such as societal biases, or in terms of intent, such as prejudice'' [\ref{itm:T5-Fairlearn}]. Other toolkits gesture towards the ``impact'' [\ref{itm:T2-ModelCards}] or ``unintended consequences'' [\ref{itm:T9-AIEthicsCards}] of systems. 

Conversely, other toolkits frame problems by articulating how AI systems can \textbf{present risks to the organizations developing or deploying them}. They highlight potential business, financial, or reputational risks, or by relating AI ethics to issues of corporate risk management more broadly. The Ethics \& Algorithms toolkit, aimed at governments and organizations who are procuring and deploying AI systems describes itself as ``A risk management framework for governments (and other people too!) to approach ethical issues.'' [\ref{itm:T7-EthicsAndAlgorithms}]. Other toolkits suggest that they can help manage business risks, in part by generating governance and compliance reports.  
In contrast with the language of harms, which focuses on people who are affected by AI systems (often by acknowledging historical harms that different groups have experienced), the language of risk is more forward facing, focusing on the potential for something to go wrong and how it might affect the organization developing or deploying the AI system---leading the organization to try to find ways to prepare contingencies for the possible negative futures it can foresee for itself. 

Not all toolkits frame AI ethics as avoiding negative outcomes, however. The integrate.ai guide uses the term ``opportunity,'' framing AI ethics in terms of \textbf{pursuing positive opportunities or outcomes}. The guide argues that AI ethics can be part of initiatives ``incentivizing risk professionals to act for quick business wins and showing business leaders why fairness and transparency are good for business'' [\ref{itm:T16-ResponsibleAI}]. The IDEO AI Ethics cards (which in some sections also frames AI ethics in terms of harms to people) also discusses capturing positive potential, writing: ``In order to have a truly positive impact, AI-powered technologies must be grounded in human needs and work to extend and enhance our capabilities, not replace them'' [\ref{itm:T9-AIEthicsCards}]. 
In these examples, AI ethics is framed as a way for businesses or the impact constituency to capture ``upside'' benefits of technology through design, development, use, and business practices. 

Some toolkits imagine that the positive or negative impacts of AI technologies will occur at a \textbf{global scale}. This is evidenced by statements such as: ``your [technology builders'] work is global. Designing AI to be trustworthy requires creating solutions that reflect ethical principles deeply rooted in important and timeless values.''  [\ref{itm:T28-HarmsModeling}]; or ``Data systems and algorithms can be deployed at unprecedented scale and speed—and unintended consequences will affect people with that same scale and speed'' [\ref{itm:T9-AIEthicsCards}].
Framing ethics globally perhaps draws attention to potential non-obvious harms or risks that might occur, prompting toolkit users to consider broader and more diverse populations who interact with AI systems. At the same time, the language of AI ethics operating at a global scale---and thus addressable at a global scale---also suggests a shared universal definition of social values, or suggests that social values have universally shared or similar impacts. This view of values as a stable, universal phenomenon has been critiqued by a range of scholars who discuss how social values are experienced in different ways, and are situated in local contexts and practices \cite{LeDantec2009Values,Houston2016Values,JafariNaimi2015ValuesHypotheses,Shilton2014HowToSeeValues,sambasivan2021reimagining,madaio2022assessing}.

\subsubsection{Sources of Legitimacy for Ethical Action} 
Toolkits' use of language also claims authority from existing discourses about what constitutes an ethical problem and how problems should be addressed. These claims help connect the toolkits’ practices to a broader set of practices or frameworks that may be more widely accepted or understood, helping to legitimize the toolkits’ perspectives and practices, and
providing a useful tactical alignment between the toolkit and existing organizational practices and resources.

Perhaps surprisingly, almost none of the toolkits provide an explicit discussion of philosophical ethical frameworks. (Although toolkits may \textit{implicitly} draw on different ethical theories, our focus in this analysis is on the explicit theories, discourses, and frameworks that are referred to in the text of the toolkits and their supporting documentation). One exception to this is the Design Ethically toolkit, which provides a brief overview of deontological ethics and consequentialism, calling them ``duty-based'' and ``results-based'' [\ref{itm:T1-EthicsKit}]. 

Several toolkits adopt the language of \textbf{``responsible innovation.''}
The Consequence Scanning toolkit was developed in the U.K. and calls itself ``an Agile event for Responsible Innovators'' [\ref{itm:T8-ConsequenceScanning}]. The integrate.ai toolkit is titled ``Responsible AI in Consumer Enterprise'' [\ref{itm:T16-ResponsibleAI}]. Fairlearn notes that its community consists of ``responsible AI enthusiasts'' [\ref{itm:T5-Fairlearn}]. Several toolkits in our corpus are listed as part of Microsoft’s ``responsible AI'' resources [\ref{T24-HAX}, \ref{itm:T27-CommunityJury}, \ref{itm:T28-HarmsModeling}]. There seems to be rhetorical power in aligning these toolkits with practices of responsible innovation, although questions about what people or groups the companies or toolkit users are responsible \textit{to} are not explicitly discussed. More broadly, what it means to align toolkits with responsible innovation is itself an open question.\footnote{With origins in the rise of science and technology as a vector of political power in the 20th century \cite{STILGOE20131568}, ``responsible innovation'' frames free enterprise as the agents of ethics, implicitly removing from frame policymakers, regulation, and other forms of popular governance or oversight. Future work should investigate more deeply what discursive work ``responsible innovation'' does in the context of AI ethics more broadly, particularly as it concerns private enterprise.}

Other toolkits look to external \textbf{laws and standards} as a legitimate basis for action; ethics is thus conceptualized as complying and acting in accordance with the law. Audit-AI, a tool that measures discriminatory patterns in data and machine learning predictions, explicitly cites U.S. labor regulations set by the Equal Employment Opportunity Commission (EEOC), writing that ``According to the Uniform Guidelines on Employee Selection Procedures (UGESP; EEOC et al., 1978), all assessment tools should comply to fair standard of treatment for all protected groups'' [\ref{itm:T19-AuditAI}]. Audit-AI similarly draws on EEOC practices when choosing a \textit{p}-value for statistical significance and choosing other metrics to define bias. This aligns the toolkit with a regulatory authority’s practices as the basis for ethics; however, it does not explicitly question whether this particular definition of fairness is applicable in contexts beyond the cultural and legal U.S. employment context \cite[cf.][]{watkins2022four}.\looseness=-1 

Several toolkits frame ethics as upholding \textbf{human rights principles}, drawing on the UN Declaration of Human Rights. In our dataset, this occurred most prominently in Microsoft’s Harms Modeling Toolkit: ``As a part of our company's dedication to the protection of human rights, Microsoft forged a partnership with important stakeholders outside of our industry, including the United Nations (UN)'' [\ref{itm:T28-HarmsModeling}].
Supported by the UN's Guiding Principles on Business and Human Rights \cite{UnitedNationsHumanRights2011}, many large technology companies have made commitments to upholding and promoting human rights.\footnote{It has been argued that involving businesses in the human rights agenda can provide legitimacy and disseminate human rights norms in broader ways than nation states could alone \cite{Ruggie2017SocialConstructionUN}. However, more recent research and commentary has been critical of technology companies' commitments to human rights \cite{Greene2019betterNicer}, with a 2019 UN report stating that big technology companies ``operate in an almost human rights-free zone.'' \cite{Alston2019UNReportPoverty}} 
This corresponds with prior research that shows how human rights discourses provide one source of values for AI ethics guidelines more broadly \cite{Jobin:2019bw}.  Many companies have existing resources or practices around human rights, such as human rights impact assessments \cite{metcalf2021algorithmicImpactAssessments, kemp2013humanRights}. Framing AI ethics as a human rights issue may help tactically align the toolkit with these pre-existing initiatives and practices.\looseness=-1


\subsection{The envisioned users and other stakeholders for toolkits (RQ2)}
\label{stakeholders}

This section asks, \textit{who is to do the work of AI ethics?} The design and supporting documentation of toolkits presupposes a particular audience---or, as \citet{mattern_2021} describes it, they ``summon'' particular users through the types of shared understanding, background knowledge, and expertise they draw on and presume their users to have.
The toolkits in our corpus mention several specific job categories \textit{internal} to the organizations in question: software engineers; data scientists; members of cross-functional or cross-disciplinary teams; risk or internal governance teams; C-level executives; board members. To a lesser extent, they mention designers. All of these categories of stakeholders pre-configure specific logics of labor and power in technology design. Toolkits that mention engineering and data science roles focus on ethics as the practical, humdrum work of creating engineering specifications and then meeting those specifications. (One toolkit, Deon, is a command-line utility for generating ``ethics checklists'') [\ref{itm:T12-Deon}]. 
For C-level executives and board members, toolkits frame ethics as both a business risk and a strategic differentiator in a crowded market. As the integrate.ai Responsible AI guide states,
``Sustainable innovation means incentivizing risk professionals to act for quick business wins and showing business leaders why fairness and transparency are good for business.'' [\ref{itm:T16-ResponsibleAI}]

Of course, stakeholders involved in AI design and development always already have their roles pre-configured by their job titles and organizational positionality; roles that the toolkits invoke and summon in their description of potential toolkit users and other relevant stakeholders. They (for example, ``business leaders'') are sensitized toward particular facets of ethics, which are made relevant to them through legible terms (for example, ``risk'').
As such, the nature of these internal (i.e., internal to the institutions developing AI) stakeholders' participation in the work of ethics is bound to vary. On what terms do these internal stakeholders get to participate? Borrowing from \citet{Hoffmann2020terms} who in turn channels \citet{ahmed2012being}, what are the ``terms of inclusion'' for each of these internal stakeholders? 


Technically-oriented tooling (like Google's What If tool [\ref{itm:T10-WhatIf}]) envisions technical staff who contribute directly to production codebases. Although toolkits rarely address the organizational positioning of engineers (and their concerns) directly, they are specific about the mechanism of action and means of participation for these technical tools. One runs statistical tests, provides assurances around edge cases, and keeps track of statistical markers like disparate impact or the \textit{p\%} rule.\looseness=-1 

For social and human-centered practices, the terms of participation are less clear. The rhetoric of these toolkits \textit{is} one of participation---between cross-functional teams (comprised of different roles), between C-suite executives and tech labor, and between stakeholders both internal and external to the organization. But no toolkit quite specifies how this engagement should be enacted. Methodological detail is scant, let alone acknowledgements of power differentials between workers and executives, or tech workers and external stakeholders. Even those rare toolkits that do acknowledge power as a factor---for example, what the Ethics \& Algorithms toolkit lists as its ``mitigation \#1''---under-specify how this power should be dealt with.

\begin{quote}
    ``Mitigation 1. Effective community engagement is people-centered, partnerships-driven, and power-aware. Engagement with the community should be social (using existing social networks and connections), technical (skills, tools, and digital spaces), physical (commons), and on equal terms (aware of and accounting for power).''
    [\ref{itm:T1-EthicsKit}]
\end{quote}

Although this ``mitigation'' refers specifically to the need to be aware of power, to account for power, it offers no specific strategies to become aware, to do such ``accounting.'' Who does that work, and how?

This question brings us to the second broad category of stakeholders invoked by toolkits---stakeholders \textit{external} to companies, described as ``the community'' above. This group variously includes
clients, vendors, customers, users, civil society groups, journalists, advocacy groups, community members, and others impacted by AI systems. These stakeholders are imagined as outside the organization in question,
sometimes by several degrees (although some, such as customers, clients, and vendors, may be variously entangled with the organization's operations \cite[cf.][]{gray2019ghost}). For example, the Harms Modeling toolkit lists ``non-customer stakeholders; direct and indirect stakeholders; marginalized populations'' [\ref{itm:T28-HarmsModeling}].
The Community Jury mentions ``direct and indirect stakeholders impacted by the technology, representative of the diverse community in which the technology will be deployed'' [\ref{itm:T27-CommunityJury}].
Google's Model Cards describes its artifacts as being for ``everyone... experts and non-experts alike'' [\ref{itm:T2-ModelCards}].
None of those toolkits, however, provide guidance on how to identify specific stakeholders \cite[cf.][]{madaio2022assessing}, or how to engage with them once they have been identified.
Indeed, the work these external stakeholders are imagined to \emph{do} in these circumstances is under-specified. Their specific roles are under-imagined, relegated to the vague ``raising concerns'' or ``providing input'' from ``on-the-ground perspectives.'' We return to this point in the following section.

\subsection{Work practices envisioned by toolkits (RQ3)}

Much of the work of ethics as imagined by the toolkits focuses on technical work with ML models, in specific workflows and tooling suites, despite claims that fairness is sociotechnical (e.g., [\ref{itm:T5-Fairlearn}]). Many toolkits aimed at design and development teams call for engagement with stakeholders external to the team or company---and for such stakeholders to inform the team about potential ethical impacts, or for the AI design team to inform and communicate about ethical risks to stakeholders. 
However, there is little guidance provided by the tools on how to do this; these imagined roles for stakeholders beyond the development team are framed as informants or as recipients of information (without the ability to shape systems’ designs) \cite[cf.][]{delgado2021stakeholder,sloane2020participation}. Moreover, the technical orientation of many toolkits may preclude meaningful participation by non-technical stakeholders. As framed by the toolkits, the work of ethics is often imagined to be done by individual data scientists or ML teams, both of whom are imagined to have the power to influence key design decisions, without considering how organizational power dynamics may shape those processes \cite[cf.][]{madaio2020co,rakova2021responsible}. The imagined work of ethics here is largely individual self-reflection, or team discussions, but without a theory of change for how self-reflection or discussions might lead to meaningful organizational shifts. 

\subsubsection{Emphasis on technical work}
Much of the work of ethics as imagined by the toolkits (and their designers) is focused on technical work with ML models, ML workflows, and ML tooling suites---with few exceptions, i.e., the Algorithmic Equity Toolkit [\ref{itm:T17-AEKit}] and others [\ref{itm:T8-ConsequenceScanning}, \ref{itm:T27-CommunityJury}] (the forms of non-technical work that these few toolkits suggest is an area for further exploration, which we discuss in Section \ref{section:recommendations-design}). This is in spite of the claims from some toolkits that ``fairness is a sociotechnical problem'' [\ref{itm:T5-Fairlearn}, \ref{itm:T27-CommunityJury}]. In practice, this means that tools’ imagined (and suggested) uses are oriented around the ML lifecycle, often integrated into specific ML tool pipelines. For instance, Amazon’s SageMaker describes how it provides the ability to ``measure biases that can occur during each stage of the ML lifecycle (data collection, model training and tuning, and monitoring of ML models deployed for inference)'' [\ref{itm:T22-SageMaker}]. Other toolkits go further, and are specifically designed to be implemented into particular ML programming tooling suites, such as Scala or Spark [\ref{itm:T18-LiFT}], TensorFlow, or Google Cloud AI platform [\ref{itm:T10-WhatIf}, \ref{itm:T20-TensorFlow}]. Some toolkits, albeit substantially fewer, provide recommendations for how toolkit users might make different choices about how to use the tool depending on where they are in their ML lifecycle [\ref{itm:T3-AIF360}]. 

However, this emphasis on technical functionality offered by the toolkits, as well as the fact that many are designed to fit into ML modeling workflows and tooling suites suggests that non-technical stakeholders (whether they are non-technical workers involved in the design of AI systems, or stakeholders external to technology companies) may have difficulty using these toolkits to contribute to the work of ethical AI. At the very least, it implies that the intended users must have sufficient technical knowledge to understand how they would use the toolkit in their work---and further reinforces that the work of AI ethics is technical in nature, despite claims to the contrary [\ref{itm:T5-Fairlearn}, \ref{itm:T27-CommunityJury}]. In this envisioned work, what role is there for designers and user researchers, for domain experts, or for people impacted by AI systems, in doing the work of AI ethics?

\subsubsection{Calls to engage stakeholders, but little guidance on how}
One of the key elements of AI ethics work suggested by toolkits involves engaging stakeholders external to the development team or their company (as discussed in Sec. \ref{stakeholders}). However, many toolkits lacked specific resources or approaches for how to do this engagement work. Toolkits often advocated for working with diverse groups of stakeholders to inform the development team about potential impacts of their systems, or to ``seek more information from stakeholders that you identified as potentially experiencing harm'' [\ref{itm:T28-HarmsModeling}]. For some toolkits, this was envisioned to take the form of user research, recommending that teams ``bring on a neutral user researcher to ensure everyone is heard'' [\ref{itm:T27-CommunityJury}] (what it means for a researcher to be ``neutral'' is left to the imagination), or to ``help teams think through how people may interact with a design'' [\ref{itm:T9-AIEthicsCards}]. Others envisioned this information gathering as workshop sessions or discussions, as in the consequence scanning guide [\ref{itm:T8-ConsequenceScanning}] or community jury approach [\ref{itm:T27-CommunityJury}].\looseness=-1

Although some toolkits called for AI development teams to learn about the impacts of their systems from external stakeholders, a smaller subset were designed to support external stakeholders or groups in better understanding the impacts of AI. For instance, the Algorithmic Equity Toolkit was designed to help citizens and community groups ``find out more about a specific automated decision system'' by providing a set of questions for people to ask to policymakers and technology vendors [\ref{itm:T17-AEKit}]. In addition, some developer-facing tools such as Model Cards were designed to provide information to ``help advocacy groups better understand the impact of AI on their communities'' [\ref{itm:T2-ModelCards}]. 

Despite these calls for engagement, toolkits lack concrete resources for precisely how to engage external stakeholders in either understanding the ethical impact of AI systems or involving them in the process of their design to support more ethical outcomes. Some toolkits explicitly name particular activities that would benefit from involving a wide range of stakeholders, such as the Harms Modeling toolkit: ``You can complete this ideation activity individually, but ideally it is conducted as collaboration between developers, data scientists, designers, user researcher, business decision-makers, and other disciplines that are involved in building the technology'' [\ref{itm:T28-HarmsModeling}].
The stakeholders named by the Harms Modeling toolkit, however, are still ``disciplines involved in building the technology'' [\ref{itm:T28-HarmsModeling}] and not, for instance, people who are harmed or otherwise impacted by the system outside of the company. Others, such as the Ethics \& Algorithms toolkit, broaden the scope, recommending that ``you will almost certainly need additional people to help - whether they are stakeholders, data analysts, information technology professionals, or representatives from a vendor that you are working with'' [\ref{itm:T7-EthicsAndAlgorithms}]. However, despite framing the activity as a ``collaboration'' [\ref{itm:T28-HarmsModeling}] or ``help'' [\ref{itm:T7-EthicsAndAlgorithms}] such toolkits provide little guidance for how to navigate the power dynamics or organizational politics involved in convening a diverse group to use the toolkit. 

\subsubsection{Theories of change}
Ethical AI toolkits present different theories of change for how practitioners using the toolkits may effect change in the design, development, or deployment of AI/ML systems. 
For many toolkits, individuals within the organization 
are envisioned to be the catalysts for change
via oaths [\ref{itm:T13-DesignEthically}] or ``an individual exercise'' [\ref{itm:T1-EthicsKit}] where individuals are prompted to ``facilitat[e] your own reflective process'' [\ref{itm:T1-EthicsKit}]. This approach is aligned with what  Boyd and others have referred to as developing ethical sensitivity \cite{boyd2020ethical,weaver2008ethical}. Some toolkits explicitly articulated the belief that individual practitioners who are aware of possible ethical issues may be able to change the direction of the design process. For instance, ``The goal of Deon is to push that conversation forward and provide concrete, actionable reminders to the developers that have influence over how data science gets done'' [\ref{itm:T12-Deon}]. However, this belief that individual data scientists ``have influence over how data science gets done'' may be at odds with the reality of organizational power structures that may lead to changes in AI design \cite[cf.][]{rakova2021responsible}.  
 
In other cases, the implicit theory of change involves product and development teams having conversations, which are then thought to lead to changes in design decisions towards more ethical design processes or outcomes. Some toolkits propose activities designed to ``elicit conversation and encourage risk evaluation as a team'' [\ref{itm:T7-EthicsAndAlgorithms}]. Others start with individual ethical sensitivity, then move to team-level discussions, suggesting that the toolkit should ``provoke discussion among good-faith actors who take their ethical responsibilities seriously'' [\ref{itm:T12-Deon}]. Such group-level activities rely on having discussions with ``good-faith actors,'' presumably those who have developed some level of individual sensitivity to ethical issues. As one toolkit suggests for these group-level conversations, ``There is a good chance someone else is having similar thoughts and these conversations will help align the team'' [\ref{itm:T9-AIEthicsCards}]. In this framing, the work of ethics involves finding like-minded individuals and getting to alignment within the team. However, this approach relies on the \textit{possibility} of reaching alignment. As such, it may not provide sufficient support for individuals whose ethical views about AI may differ from their team. Individuals may feel social pressure from others on their team to stay silent, or not appear to be contrarian in the face of consensus from the rest of their team \cite[cf.][]{madaio2020co}.

In fact, despite many toolkits' claims to empower individual practitioners to raise issues, toolkits largely appeared not to address fundamental questions of worker power and collective action. For instance, the IDEO AI Ethics Cards state that ``all team members should be empowered to trust their instincts and raise this Pause flag… at any point if a concept or feature does not feel human-centered'' [\ref{itm:T9-AIEthicsCards}], and similarly the Design Ethically Toolkit advises that ``Having a variety of different thinkers who are all empowered to speak in the brainstorm session makes a world of a difference'' [\ref{itm:T13-DesignEthically}]. However, the Design Ethically toolkit was the only example in our corpus that provided resources to support workplace organizing to meaningfully secure power for tech workers in driving change within their organizations. 

Finally, other toolkits pose theories of change that suggest that pressure from external sources (i.e., media, public pressure or advocacy, or other civil society actors or organizations) may lead to changes in AI design and deployment (usually implied to be within corporate or government contexts). The Algorithmic Equity Kit in particular, is explicitly designed to provide resources for ``community groups involved in advocacy campaigns'' [\ref{itm:T17-AEKit}] to help support that advocacy work. Other toolkits, such as the Ethics \& Algorithms Toolkit, focus on government agencies using AI that are ``facing increasing pressure from the public, the media, and academic institutions to be more transparent and accountable about their use'' [\ref{itm:T7-EthicsAndAlgorithms}]. As such, the toolkit offers resources for government agencies to respond to such pressure and provide more transparency and accountability in their algorithmic systems.
 
More generally, many toolkits enact some form of solutionism---the belief that ethical issues that may arise in AI design can be solved with the right tool or process (typically the approach they propose). 
Some tools [e.g., \ref{itm:T2-ModelCards}, \ref{itm:T3-AIF360}, \ref{itm:T10-WhatIf}, \ref{itm:T20-TensorFlow}] suggest that ethical values such as fairness can be achieved via technical tools alone:
``If all fairness metrics are fair, The Bias Report will evaluate the current model as fair.'' [\ref{itm:T6-Aequitas}].
Some toolkits (albeit fewer) do note the limitations of purely technical solutions to fundamentally sociotechnical problems [\ref{itm:T3-AIF360}, \ref{itm:T5-Fairlearn}, \ref{itm:T10-WhatIf}], as in AIF360’s documentation, which states that ``the metrics and algorithms in AIF360… clearly do not capture the full scope of fairness in all situations'' [\ref{itm:T3-AIF360}]. As the What-If tool documentation states, ``There is no one right [definition of fairness], but we probably can agree that humans, not computers, are the ones who should answer this question'' [\ref{itm:T10-WhatIf}]. 
However, even with these acknowledgements, the documentation goes on to note the important role that the toolkit plays in enabling humans to answer that question, as ``What-If lets us play `what if' with theories of fairness, see the trade-offs, and make the difficult decisions that only humans can make'' [\ref{itm:T10-WhatIf}].
 
These general framings suggest a particular flavor of solutionism, in which the work 
of ethics in AI design involves following a particular process (i.e., the one proposed by the toolkit). 
Toolkits propose ethical work practices that fit into existing development processes [e.g., \ref{itm:T12-Deon}], in ways that suggest that all that is needed is the addition of an activity or discussion prompt and not, for instance, fundamental changes to the corporate values systems or business models that may lead to harms from AI systems. Some toolkits were explicit that ethical AI work should not significantly disrupt existing corporate priorities, saying, ``Business goals and ethics checks should guide technical choices; technical feasibility should influence scope and priorities; executives should set the right incentives and arbitrate stalemates'' [\ref{itm:T16-ResponsibleAI}].\looseness=-1 

\section{Discussion}

Throughout these toolkits, we observed a mismatch between the imagined roles and work practices for ethics in AI and the support the toolkits provided for achieving those roles and practices. Specifically, despite rhetoric from the documentation of many toolkits that the work of ethics is \textit{socio}technical, involving contributions from a variety of stakeholders, the actual design and functionality of the majority of toolkits involved \textit{technical} work for primarily developers and data scientists. Toolkits suggested multi-stakeholder approaches to addressing ethical issues in sociotechnical ways, but most toolkits provided little scaffolding for the social dimensions of ethics or for engaging stakeholders from multiple (non-technical) backgrounds. These technosolutionist approaches to AI ethics suggest that AI ethics toolkits may act as a ``technology of de-politicization'' \cite[cf.][]{hitzig2020normative}, sublimating sociopolitical considerations in favor of technical fixes. With few exceptions [e.g., \ref{itm:T17-AEKit}], the toolkits took a decontextualized approach to ethics, largely divorced from the sociopolitical nuance of what ethics might mean in the contexts in which AI systems may be deployed, or how ethical work practices might be enacted within the organizational contexts of the sites of AI production (e.g., technology companies). In such a decontextualized view of ethics, toolkit designers envision individual users who have the agency to make decisions about their design of AI systems, and who are not beholden to the role of power dynamics within the workplace: organizational hierarchies, misaligned priorities, and incentives for ethical work practices---key considerations for the use of AI ethics toolkits, given the reality of business priorities and profit motives.\looseness=-1 

When toolkits \textit{did} attend to how ethical work might fit within business processes, many of them leveraged discourses of business risk and responsible innovation to help motivate adoption of ethics tools and processes. These discourses may function tactically \cite[cf.][]{wong2021tactics} as a way to allow toolkits to tap into existing institutional processes and resources they may not otherwise have access to (for example, mechanisms for managing legal liability). However, in so doing, companies may sidestep questions of how logics of capital accumulation themselves shape the capacity for AI systems to exert harms and shape the sociotechnical imaginaries \cite[cf.][]{jasanoff2015dreamscapes} for what ethics might mean---or foreclose alternative ways of conceptualizing ethics. As a result, ethical concerns may be sublimated to the interests of capital. In the following sections, we unpack implications of our findings for AI ethics toolkit researchers and designers.\looseness=-1

\subsection{Reflections and Implications for Research} 
As the prior sections suggest, the content and guidance provided by toolkits, as well as the metaphor and format of ``toolkits'' as a predominant way to address AI ethics, constructs particular ways of seeing the world---what constitutes an ethical problem, who should be responsible for addressing those problems, and what are the legitimate practices for addressing them. We underscore this point by using the metaphor of ``seeing like a toolkit,'' to draw attention to two ideas. 

First, although toolkits provide a useful format for sharing information and practices across boundaries and contexts, an over-reliance on toolkits may risk decontextualizing or abstracting away from the social and political contexts where AI systems are deployed and governed, and from the organizational contexts in which those toolkits may be used. Toolkits, by design, are intended to be portable objects usable across a variety of contexts \cite{mattern_2021, kelty_2018}---but as a result, ethical AI toolkits may act as a ``device for decontextualizing'' \cite{kelty_2018}. This portability may allow toolkits to be more generalizable or scalable by ``mediating between the local and the universal'' \cite{mattern_2021} in order to support their adoption and use across multiple contexts. However, the flattening of local distinctiveness in order to be more easily transportable across contexts \cite{redfield2013life} brings with it particular risks for ethical AI. As Selbst et al., have written, efforts for fairness in AI run the risk of what they have referred to as ``abstraction traps,'' or abstracting away crucial elements of the social context in which AI systems are deployed and within which fairness and ethical considerations must be understood \cite{selbst2019fairness}. As a result, toolkits that are explicitly designed to be decontextualized---both from the social context where AI systems will be deployed (and within which ethics must be understood \cite{sambasivan2021reimagining}) and from the organizational context in which those toolkits may be used \cite{suchman2002located}---may inadvertently suggest to their users that either the context does not matter for the work of ethics, or that it is up to the toolkit user to do the work of \textit{re}contextualizing, or translating its methods for their context of use and deployment (cf. \cite{morley2021initial}). However, this is quite a burden for the toolkits to place on their users, particularly as the imagined users of many ethical AI toolkits appear to be largely technical practitioners who may not have the training or background to do such contextualization and translation work.

This pattern of decontextualization of toolkits mirrors Scott's concepts of legibility and simplification in statecraft.\footnote{whose book \textit{Seeing Like a State} informs the title of this paper} In order to govern, the state employs techniques such as standardized measurement or systems of private property ownership to make local heterogeneous practices legible, but this also serves to simplify and standardize understandings of social practices which may not equate with local experiences \cite{Scott1998seeing}. Similarly, for toolkits to be legible among communities of practice and organizational structures that seek to build systems at scale, toolkits make ethical practices legible in ways that are often simplified and do not account for the hetereogeneity of contextual experiences and on the ground practices of doing AI ethics, requiring users who can do this difficult translation work.

Second, these toolkits represent a form of ``professional vision'' that may inadvertently promote a solutionist orientation to AI ethics. As Goodwin has argued, ``professional vision'' is how the discursive practices of professional cultures shape how we see the world in socially situated and historically constituted ways \cite{goodwin2015professional}. Similarly, in Silbey's work on industrial safety culture, she argues that disasters that are not spectacular or sudden---such as slow-acting oil leaks---are often ignored, ``existing physically, but not in any organizationally cognizable form'' \cite{silbey2009taming}. For ethics in AI, the discursive practices instantiated in our tools shape how the field sees the ethical terrain for action---what are the objects of concern, how might they be made legible or amenable to action, what resources might be marshalled to address them, and by whom. Likewise, problems left outside of toolkits' purview may risk not being seen as legitimate ethical issues by practitioners.

The tools curated within a toolkit are intended to solve particular problems (here, problems related to the ethics of AI), but the metaphor of the toolkit itself may reinforce a solutionist framing, suggesting to their users that ethical problems can in fact, be \textit{solved} by using the tools or processes therein---for instance, that AI systems can be ``de-biased,'' which they cannot be \cite{blodgett2020language,hoffmann2019wherefairness}---rather than mitigating their potential for harm. This solutionist orientation is not limited to toolkits; indeed, Selbst et al. have written about the solutionist trap for fairness in sociotechnical systems more generally \cite{selbst2019fairness}, but the genre of the toolkit may inadvertently reinforce the idea of ethics as a managerial exercise \cite{kelty_2018}, or a technical solution to fundamentally contextual and contested challenges (cf. \cite{selbst2019fairness, stark2021critical}). As a result, this framing may inhibit investment (of time, attention, resources) into alternative approaches that do not fit within the confines of the solutionist orientation of a toolkit, or foreclose alternative theories of change (such as a focus on the political economy of AI development \cite{stark2021critical}). This may also lead to false expectations (from practitioners using the toolkit as well as stakeholders and communities impacted by AI), potentially leading to frustration, resentment, and further harm when those expectations for solved problems are not met. Others have discussed how corporate dicourse of ``solving'' ethical issues are often rooted in public relations goals or economic self-interest \cite{mcmillan2019againstethical,bietti2020ethicswashing}. 

This is a broader issue for the field. Metcalf and Moss discuss how ethics in Silicon Valley is in part framed through the lenses of technological solutionism and market fundamentalism---that an optimal set of tools, procedures, or criteria will lead to an ethical outcome, and that ethical solutions should be pursued within the boundaries of what the market finds profitable \cite{Metcalf2019OwningEthics}. These lenses miss out on the value of non-technical expertise and practices, as well as a broader array of potential ethical (if less profitable) alternatives. What do we lose when we fail to grapple with capital as a force in shaping the ethical considerations of AI? We note that these critiques are not a call to abandon toolkits altogether, but rather an interrogation of what politics we might (unintentionally) embed when framing an AI ethics intervention as a ``toolkit.'' What are the political choices one makes when one creates a toolkit, and how can we make those choices more intentional? Although we find that AI ethics toolkits tend to focus on technical practices in ways that may be decontextualized from the wider social and political context, we are inspired by toolkits in other domains that explicitly engage in questions of politics and power, for example toolkits that serve as methods of participatory engagement to purposefully include broader communities to consider issues of justice \cite[e.g.,][]{mattern_2021,bray2022radical}.



We also consider the politics of the choice of deciding to make a ``toolkit'' versus making something else. We thus ask what ways of ``seeing'' AI ethics do \textit{all} toolkits miss? What are new ways of seeing that can produce new, practical interventions? New approaches might move beyond toolkits and look to other theories of change, such as political economy \cite{stark2021critical}.  
However, we as authors note that our situatedness in particular debates in the West may occlude our sensitivity to alternative ethical frameworks. 
Indigenous notions of ``making kin'' \cite{lewis2018making} could reveal radical new possibilities for what AI ethics could be, and by what processes it may be enacted.
How can we, as a research community, make space for such alternatives?
Following from this problem-posing orientation, we do not offer solutions here, but instead pose these as questions for researchers, practitioners, and communities to address through developing alternatives to the dominant paradigm of the toolkit. Some promising examples include the People's Guide to AI zine \cite{Onuoha2018PeoplesGuideAI}; 
J. Khadijah Abdurahman's and We Be Imagining's call for lighting ``alternate beacons'' to help ``organize for different futures'' for technology development \cite{Abdurahman2021Body}; 
and the AI Now Institute's series on a new lexicon to offer narratives beyond those from the Global North to critically study AI  \cite{Raval2021NewAILexicon}, among others \cite[e.g.,][]{bray2022radical}. 
We call on the CSCW community and others (e.g., FAccT, CHI) to amplify and expand these efforts.\looseness=-1 

\subsection{Recommendations for Toolkit Design}
\label{section:recommendations-design}
Practitioners will continue to require support in enacting ethics in AI, and toolkits are one potential approach to provide such support, as evidenced by their ongoing popularity. Although much of this paper has focused on a critical analysis of toolkits, we offer suggestions for toolkit design following the ``practical turn'' in values in design research \cite[pg9]{flanagan2014values}---i.e., if we accept that toolkits can embody and promote particular social values, we might consider an additional (or alternative) set of values in the design of toolkits. We acknowledge that toolkits alone will not solve all the problems of addressing AI ethics, but they can nevertheless be improved to better consider the social and organizational contexts where they might be deployed. 

Our findings suggest three concrete recommendations for improving toolkits' potential to support the work of AI ethics. Toolkits should: (1) provide support for the non-technical dimensions of AI ethics work; (2) support the work of engaging with stakeholders from non-technical backgrounds; (3) structure the work of AI ethics as a problem for collective action. 



\subsubsection{Embrace the non-technical dimensions of ethics work}
Despite emerging awareness that fairness is \textit{socio}technical, the majority of toolkits provided resources to support technical work practices (although some toolkits called for their users to engage in other forms of work [e.g., \ref{itm:T5-Fairlearn}]). This might entail resources to support understanding the theories and concepts of ethics in non-technical ways,\footnote{Note that Fairlearn [\ref{itm:T5-Fairlearn}] has---since we conducted the data analysis for this paper---published resources in its user guide for understanding social science concepts such as construct validity for concepts such as fairness \cite{jacobs2021measurement} and explanations of sociotechnical abstraction traps \cite{selbst2019fairness}.} as well as resources drawing from the social sciences for understanding stakeholders' situated experiences and perceptions of AI systems and their impacts. 
For instance, toolkit designers might incorporate methods from qualitative research, user research, or value-sensitive design (e.g., \cite{friedman2002value}), as some existing tools suggest (e.g., [\ref{itm:T27-CommunityJury}]). Although some AI ethics education tools are beginning to be designed with these perspectives (e.g., value cards \cite{shen2021valuecards}), fewer practitioner-oriented toolkits utilize them. As a precursor to this, practitioners may need support in identifying the stakeholders for their systems and use cases \cite[cf.][]{madaio2022assessing}, in the contexts in which those systems are (or will be) deployed, including community members, data subjects, or others beyond the users, paying customers, or operators of a given AI system. Approaches such as stakeholder mapping from fields like Human-Computer Interaction \cite[e.g.,][]{yoo2018stakeholder} may be useful here, and such resources may be incorporated into AI ethics toolkits.

\subsubsection{Support for engaging with stakeholders from non-technical backgrounds}
Although many toolkits call for engaging stakeholders from different backgrounds and with different forms of expertise (internal stakeholders such as designers or business leaders; external stakeholders such as advocacy groups and policymakers), the toolkits themselves offer little support for how their users might bridge such disciplinary divides, further contributing to the mismatch between the rhetorical promise of toolkits and their current design. 
Toolkits should thus support this translational work.\footnote{Some emerging work is exploring the role of ``boundary objects'' \cite[cf.][]{star1989structure} to help practitioners align on key concepts and develop a shared language, e.g., \hyperlink{https://events.withgoogle.com/pair-symposium-2020/}{PAIR Symposium 2020}, although this work has not focused on ethics of AI specifically.}
This might entail, for instance, asking what fairness means to the various stakeholders implicated in ethical AI, or communicating the output of algorithmic impact assessments (e.g., various fairness metrics) in ways that non-technical stakeholders can understand and work with \cite{cheng2021soliciting, shen2020designing}. The Algorithmic Equity Toolkit (whose design process is discussed in \cite{10.1145/3442188.3445938}) tackles this challenge from the perspective of community members and groups, providing resources to these external stakeholders to support their advocacy work [\ref{itm:T17-AEKit}]. Meanwhile, recent research has explored how to engage non-technical stakeholders in discussions about tradeoffs in model performance \cite[e.g.,][]{cheng2021soliciting, shen2020designing, shen2021valuecards}, or in participatory AI design processes more generally \cite{delgado2021stakeholder,sloane2020participation}, although such approaches have largely not been incorporated into toolkits (with few recent exceptions \cite[e.g.,][]{shen2022model}). Moreover, approaches that involve stakeholders impacted by AI conducting ``crowd audits'' of algorithmic harms \cite[e.g.,][]{shen2021everyday} have not yet made their way into the toolkits we analyzed, where the results of such crowd audits might be used to shape AI practitioners' development practices.\looseness=-1 

\subsubsection{Structure the work of AI ethics as a problem for collective action}
One question we found palpably missing in the toolkits we analyzed was, \textit{how do toolkits support stakeholders in grappling with organizational dynamics involved in doing the work of ethics?} Silbey has written about the ``safety culture'' promoted in other high-stakes industries (e.g., fossil fuel extraction), where the responsibility to avoid catastrophe is too often located in the behaviors and attitudes of individual actors---typically those with the least power in the organization---rather than systemic processes or organizational oversight \cite{silbey2009taming}. To address this gap, toolkits could provide support for helping practitioners communicate to organizational leadership and advocate for the need to engage in ethical AI work practices, or advocate for additional time or resources to do this work. One form this might take is providing support for strategic alignment of ethics discourses with business priorities and discourses (e.g., business risk, responsible innovation, corporate social responsibility, etc). However, these discourses bring risks: the aims and values of ethical AI could be subverted by business priorities. For instance, \citet{madaio2022assessing} discuss how business priorities for AI deployment across market tiers may subvert practitioners' goals for fairness work.
Given the risk that such an approach might smuggle in business logics that subvert ethical aims (see Sec. \ref{discourses}), toolkit designers might instead consider how to support the users of their toolkits in becoming aware of the organizational power dynamics that may impact the work of ethics (e.g., power mapping exercises \cite{LittleSis2017MapThePower}), 
including identifying institutional levers they can pull to shape organizational norms and practices from the bottom up. In addition, toolkits should structure ethical AI as a problem for collective action for multiple groups of stakeholders, rather than work for individual practitioners. This may involve supporting collective action by workers within tech companies, or fostering communities of practice of professionals working on ethical AI across institutions (to share knowledge and best practices, as well as shift professional norms and standards), or supporting collective efforts for ethical AI across industry professionals designing AI and communities impacted by AI. This might also involve providing support for organizing collective action in the workplaces, such as unions, tactical walkouts, or other uses of labor power based on their role in technology production \cite{Khovanskaya2019dataRhetoric,wong2021tactics,stark2021critical, Ozoma2021TechWorkerHandbook}. Prior research found that technology professionals pursuing design justice sought project- and institutional-level tools and interventions rather than individual-level ones \cite{Spitzberg2020}. However, few toolkits we saw (with the Data Ethically toolkit as a notable exception [\ref{itm:T13-DesignEthically}]) provide resources to inform and support practitioners about the role of collective action in ethical AI.\looseness=-1

\subsection{Limitations and Future Work}
\label{limitations}

We examined a small subset of toolkits which may not be representative of all AI ethics toolkits. Most of the toolkits we examined were from tech companies and academia, and we may thus have missed out on toolkits developed by nonprofits, civil society, or government agencies. Furthermore, the toolkits we examined largely skewed towards industry practitioners as the envisioned users (with some exceptions; e.g., [\ref{itm:T17-AEKit}]), and were largely intended to fit into AI development processes (as suggested by the large proportion of toolkits that were open source code). As such, future work should explicitly target toolkits intended to be used by policymakers, civil society, or community stakeholders more generally. Recognizing that creating technical tools can re-inscirbe the harms they seek to address (e.g., \cite{green2021datascience,hoffmann2019wherefairness}) in addition to re-designing the politics of toolkits, future work should also investigate other forms of political action that consider and address the social and institutional aspects of technology development. 

In addition, our corpus was built from search queries; as such, searching for toolkits using terms we did not include here may result in identifying toolkits that we did not include in our corpus. More broadly, our positionality has shaped how we approach our research, including the research questions we chose, the toolkits we identified, and how we coded and interpreted our data. As Sambasivan et al. \cite{sambasivan2021reimagining} (among others, such as \citet{ding2018deciphering}) have pointed out, AI ethics may mean different things in different cultural contexts, including 
relying on different legal frameworks, and aiming towards fundamentally different outcomes. Our corpus is necessarily partial and reflective of our positionality and cultural context.\looseness=-1 

\section{Conclusion}

This paper investigates how AI ethics toolkits frame and embed particular visions for what it means to do \textit{the work of addressing ethics}. Based on our findings, we recommend that designers of AI ethics toolkits should better support the social dimensions of ethics work, provide support for engaging with diverse stakeholders, and frame AI ethics as a problem for collective action rather than individual practice. Toolkit development should be tied more closely to empirical research that studies the social, organizational, and technical work required to surface and address ethical issues. Creating tools or resources in a format that challenges the notions of the ``toolkit'' \textit{per se} may open up the design space to foster new approaches to AI ethics. Although no single artifact alone will solve all AI ethics problems, intentionally diversifying the forms of work that such artifacts envision and support may enable more effective ethical interventions in the work practices adopted by developers, designers, researchers, policymakers, and other stakeholders. 

\begin{acks}
Thank you to Emma Lurie, Zoe Kahn, Ken Holstein, our colleagues at the UC Berkeley Center for Long-Term Cybersecurity and Microsoft Research, and the anonymous reviewers for their comments and feedback on this work.
\end{acks}

\bibliographystyle{ACM-Reference-Format}
\bibliography{bibliography}

\appendix

\section{Toolkit Listing and Analysis}
\label{section:toolkit-list}

\begin{enumerate}[label=T\arabic*]
    \item \label{itm:T1-EthicsKit}Ethics Kit, \url{http://ethicskit.org/tools.html}
    \item \label{itm:T2-ModelCards}Model Cards, \url{https://modelcards.withgoogle.com/about}
    \item \label{itm:T3-AIF360} AI Fairness 360, \url{https://aif360.mybluemix.net/}
    \item \label{itm:T4-InterpretML} InterpretML, \url{https://github.com/interpretml/interpret}
    \item \label{itm:T5-Fairlearn} Fairlearn,	\url{https://fairlearn.github.io/}
    \item \label{itm:T6-Aequitas} Aequitas,	\url{http://aequitas.dssg.io/}
    \item \label{itm:T7-EthicsAndAlgorithms}\label{itm:T15-EthicsAndAlgorithms} Ethics \& Algorithms Toolkit \url{https://ethicstoolkit.ai/}
    \item \label{itm:T8-ConsequenceScanning} Consequence Scanning Kit,	\url{https://www.doteveryone.org.uk/project/consequence-scanning/}
    \item \label{itm:T9-AIEthicsCards} AI Ethics Cards, \url{https://www.ideo.com/post/ai-ethics-collaborative-activities-for-designers}
    \item \label{itm:T10-WhatIf} What If Tool, \url{https://pair-code.github.io/what-if-tool/}
    \item \label{itm:T11-DigitalImpactToolkit} Digital Impact Toolkit, \url{https://digitalimpact.io/toolkit/}
    \item \label{itm:T12-Deon}	Deon Ethics Checklist, \url{http://deon.drivendata.org/}
    \item \label{itm:T13-DesignEthically} Design Ethically Toolkit, \url{https://www.designethically.com/toolkit}
    \item \label{itm:T14-Lime} Lime, \url{https://github.com/marcotcr/lime}
    \item \label{itm:T26-WeightsBalances} Weights and Biases, \url{https://wandb.ai/site}
    \item \label{itm:T16-ResponsibleAI}	Responsible AI in Consumer Enterprise,	\url{https://static1.squarespace.com/static/5d387c126be524000116bbdb/t/5d77e37092c6df3a5151c866/1568138185862/Ethics-of-artificial-intelligence.pdf}
    \item \label{itm:T17-AEKit}	Algorithmic Equity Toolkit (AEKit),	\url{https://www.aclu-wa.org/AEKit}
    \item \label{itm:T18-LiFT} LinkedIn Fairness Toolkit (LiFT), 	\url{https://github.com/linkedin/LiFT}, \url{https://engineering.linkedin.com/blog/2020/lift-addressing-bias-in-large-scale-ai-applications}
    \item \label{itm:T19-AuditAI}	Audit AI, \url{https://github.com/pymetrics/audit-ai}
    \item \label{itm:T20-TensorFlow} TensorFlow Fairness Indicators,	\url{https://github.com/tensorflow/fairness-indicators}
    \item \label{itm:T21-JudgmentCall} Judgment Call,	\url{https://docs.microsoft.com/en-us/azure/architecture/guide/responsible-innovation/judgmentcall}
    \item \label{itm:T22-SageMaker} SageMaker Clarify, \url{https://sagemaker-examples.readthedocs.io/en/latest/sagemaker_processing/fairness_and_explainability/fairness_and_explainability.html}
    \item \label{itm:T23-NLPChecklist}	NLP CheckList, 	\url{https://github.com/marcotcr/checklist}
    \item \label{T24-HAX} HAX Workbook and Playbook, 	\url{https://www.microsoft.com/en-us/haxtoolkit/workbook/}
    \item \label{itm:T27-CommunityJury}	Community Jury, \url{https://docs.microsoft.com/en-us/azure/architecture/guide/responsible-innovation/community-jury/}
    \item \label{itm:T28-HarmsModeling}	Harms Modeling,	\url{https://docs.microsoft.com/en-us/azure/architecture/guide/responsible-innovation/harms-modeling/}
    \item \label{itm:T29-AlgorithmicAccountabilityPolicy}	Algorithmic Accountability Policy Toolkit,	\url{https://ainowinstitute.org/aap-toolkit.pdf}

\end{enumerate}


\begin{table*}[]
\small
\caption{Analyzed Dimensions of Toolkits}
\label{tab:toolkits-analysis}
\begin{tabular}{p{0.04\textwidth}p{0.1\textwidth}p{0.2\textwidth}p{0.10\textwidth}p{0.2\textwidth}p{0.2\textwidth}}

\toprule
\textbf{ID} & \textbf{Toolkit Name} & \textbf{Toolkit Author(s)} & \textbf{Author Types} & \textbf{Audience(s)} & \textbf{Form Factor} \\
\midrule
\ref{itm:T1-EthicsKit} & Ethics Kit & Open Data Institute, Common Good, Co-op Digital, Hyper Island, Plot & Non-Profit, Design Agency & Designers & Design Exercises, Worksheets \\
\midrule
\ref{itm:T2-ModelCards} & Model Cards & Google & Technology Company & Developers, Policymakers, Analysts, Advocates, Users & Examples,  Webpage \\
\midrule
\ref{itm:T3-AIF360} & AI Fairness 360 & IBM & Technology Company & Data Scientists & Open Source Code, Documentation, Code Examples, Tutorials \\
\midrule
\ref{itm:T4-InterpretML} & InterpretML & Microsoft & Technology Company & Data Scientists & Open Source Code, Documentation, Code Examples \\
\midrule
\ref{itm:T5-Fairlearn} & Fairlearn & Miro Dudik (Microsoft Research), Microsoft Research, Open Source Community & Technology Company; Open Source Community & Data Scientists & Open Source Code, Documentaiton, User Guide, Code Examples \\
\midrule
\ref{itm:T6-Aequitas} & Aequitas & University of Chicago Center for Data Science and Public Policy & University & ML Developers, Analysts, Policymakers & Open Source Code, Web Audit Tool, Example, Documentation \\
\midrule
\ref{itm:T7-EthicsAndAlgorithms} & Ethics \& Algorithms Toolkit & Johns Hopkins Center for Government Excellence (GovEx), City and County of San Francisco, Harvard DataSmart, Data Community DC & University, Government Agency, Non-Profit & Government Leaders, Stakeholders, Data Analysts, Information Technology Professionals, Vendor Representatives & Guide, Worksheets \\
\midrule
\ref{itm:T8-ConsequenceScanning} & Consequence Scanning Kit & Dot Everyone & Non-Profit & Team Members, User Advocates, Tech and Business Specialists, Business or External Stakeholders & Manual, Exercises \\
\midrule
\ref{itm:T9-AIEthicsCards} & AI Ethics Cards & IDEO & Design Agency & Designers & Cards \\
\midrule
\ref{itm:T10-WhatIf} & What If Tool & People + AI Research Team (Google) & Technology Company & Data scientists & Open Source Code, Tutorials, Documentation, Examples \\
\midrule
\ref{itm:T11-DigitalImpactToolkit} & Digital Impact Toolkit & Stanford Digital Civil Society Lab & University & Civil Society Organizations & Checklists, Worksheets, Reading Materials \\
\midrule

\end{tabular}
\end{table*}

\begin{table*}[]
\small
\begin{tabular}{p{0.04\textwidth}p{0.1\textwidth}p{0.2\textwidth}p{0.10\textwidth}p{0.2\textwidth}p{0.2\textwidth}}

\toprule
\textbf{ID} & \textbf{Toolkit Name} & \textbf{Toolkit Author(s)} & \textbf{Author Types} & \textbf{Audience(s)} & \textbf{Form Factor} \\
\midrule
\ref{itm:T12-Deon} & Deon Ethics Checklist & DrivenData & Non-Profit & Developers & Checklist, Open Source Code, Documentation \\
\midrule
\ref{itm:T13-DesignEthically} & Design Ethically Toolkit & Kat Zhou & Tech Worker & Designers & Exercises, Worksheets \\
\midrule
\ref{itm:T14-Lime} & Lime & Macro Ribeiro, Sameer Singh, Carlos Guestrin (University of Washington); Open Source Community & University; Open Source Community & Data Scientists & Open Source Code, Documentation \\
\midrule
\ref{itm:T26-WeightsBalances} & Weights and Biases & Weights and Biases & Technology Company & Developers & SaaS product, Articles \\ 
\midrule
\ref{itm:T16-ResponsibleAI} & Responsible AI in Consumer Enterprise & integrate.ai & Technology Company & Organizations, Executive Leadership, Implementation teams & Guide, Framework \\ 
\midrule
\ref{itm:T17-AEKit} & Algorithmic Equity Toolkit (AEKit) & ACLU of Washington, Critical Platform Studies Group, Tech Fairness Coalition & University; Non-Profit & Community Groups & Activities \\
\midrule
\ref{itm:T18-LiFT} & LinkedIn Fairness Toolkit (LiFT) & LinkedIn & Technology Company & Machine Learning Developers & Open Source Code, Documentation, Blog \\
\midrule
\ref{itm:T19-AuditAI} & Audit AI & Pymetrics & Technology Company & Data Scientists & Open Source Code, Documentation, Examples \\
\midrule
\ref{itm:T20-TensorFlow} & TensorFlow Fairness Indicators & Google & Technology Company & "Teams" & Open Source Code, Documentation, Examples \\
\midrule
\ref{itm:T21-JudgmentCall} & Judgment Call & Microsoft Research & Technology Company & Technology builders, managers, designers & Cards, Activities \\
\midrule
\ref{itm:T22-SageMaker} & SageMaker Clarify & Amazon & Technology Company & "AWS customers" & Proprietary Code, Documentation, Example \\
\midrule
\ref{itm:T23-NLPChecklist} & NLP CheckList & Marco Tulio Ribeiro (Microsoft Research), Tongshuang Wu (University of Washington), Carlos Guestrin (University of Washington), Smaeer Singh (UC Irvine) & University; Technology Company & Team & Open Source Code, Documentation, Examples \\
\midrule

\end{tabular}
\end{table*}

\begin{table*}[]
\small
\begin{tabular}{p{0.04\textwidth}p{0.1\textwidth}p{0.2\textwidth}p{0.10\textwidth}p{0.2\textwidth}p{0.2\textwidth}}

\toprule
\textbf{ID} & \textbf{Toolkit Name} & \textbf{Toolkit Author(s)} & \textbf{Author Types} & \textbf{Audience(s)} & \textbf{Form Factor} \\
\midrule
\ref{T24-HAX} & HAX Workbook and Playbook & Microsoft Research & Technology Company & UX, AI, project management, and engineering teams & Guide, Workbook/Worksheets, Examples, Guidelines \\
\midrule
\ref{itm:T27-CommunityJury} & Community Jury & Microsoft & Technology Company & Product Team & Activity \\
\midrule
\ref{itm:T28-HarmsModeling} & Harms Modeling & Microsoft & Technology Company & Technology Builders & Activity \\
\midrule
\ref{itm:T29-AlgorithmicAccountabilityPolicy} & Algorithmic Accountability Policy Toolkit & AI Now & Non-Profit & Legal and Policy Advocates & PDF Guide \\
\midrule

\end{tabular}
\end{table*}

\received{July 2022}
\received[revised]{October 2022}
\received[accepted]{January 2023}

\end{document}